\documentclass[journal]{IEEEtran}

\usepackage{xr-hyper}
\usepackage{hyperref}

\usepackage[acronym,nonumberlist,nogroupskip,style=super]{glossaries}
\newacronym{iot}{IoT}{Internet of Things}

\usepackage{cite}
\usepackage{amsmath,amssymb,amsfonts}
\usepackage{graphicx}
\usepackage{textcomp}
\usepackage[dvipsnames]{xcolor}

\usepackage[font=footnotesize,labelfont=bf]{caption}

\usepackage[export]{adjustbox}
\usepackage{pifont}
\usepackage{multirow}
\usepackage{tabularx}
\usepackage{algorithmicx}
\usepackage[noend]{algpseudocode}
\usepackage[T1]{fontenc}
\usepackage{algorithm}
\usepackage{subfig}
\usepackage{soul}
\usepackage{booktabs}
\usepackage{rotating}
\usepackage{siunitx,etoolbox}
\usepackage{color, colortbl}
\usepackage{comment}

\hypersetup{
    colorlinks=true,
    linkcolor=blue,
    citecolor=blue,
    urlcolor=blue
}

\begin{document}

\title{LSTM based IoT Device Identification}

\author{Kahraman Kostas \\
ORCID: 0000-0002-4696-1857
}

\maketitle

\begin{abstract}
While the use of the Internet of Things is becoming more and more
popular, many security vulnerabilities are emerging with the large
number of devices being introduced to the market. In this
environment, IoT device identification methods provide a preventive
security measure as an important factor in identifying these devices
and detecting the vulnerabilities they suffer from. In this study,
we present an end-to-end machine learning pipeline that identifies
IoT devices in the Aalto university dataset (IoT devices captures) using Long Short-Term Memory
(LSTM) networks. Raw network packet captures (PCAP) are processed
into 25 engineered features, which are then arranged as sliding-window
time-series sequences. We systematically evaluate sequence lengths
from 2 to 20, reporting that performance improves approximately
linearly up to length 6 and thereafter in a wave-like pattern,
reaching its peak at length 18. On the final held-out test set with
the optimal configuration, the model achieves an accuracy of 79.85\%
and a macro-averaged F1-score of 75.70\% across 27 device classes.
\end{abstract}

\section{Introduction}

The rapid proliferation of Internet of Things (IoT)  devices introduces an expanding
attack surface into modern networks. Unlike traditional computing
endpoints, many IoT devices lack robust built-in security mechanisms
and are often deployed without adequate monitoring. Identifying the
exact type of a device on the network is a fundamental first step
toward enforcing appropriate security policies, detecting anomalous
behaviour, and mitigating device-specific
vulnerabilities~\cite{haddadpajouh2018deep, koroniotis2019towards}.

Feed-forward artificial neural network (ANN) models such as
Convolutional Neural Networks (CNNs) are highly effective on static,
grid-structured data. However, network traffic is inherently
sequential: each packet in a flow carries temporal context from its
predecessors. This sequential structure calls for a specialised class
of neural architectures designed to capture temporal
dependencies~\cite{Nguyen_1, Nguyen_2, olah2015understanding}.

\begin{figure}[htbp]
    \centering
    \includegraphics[width=0.7\linewidth]{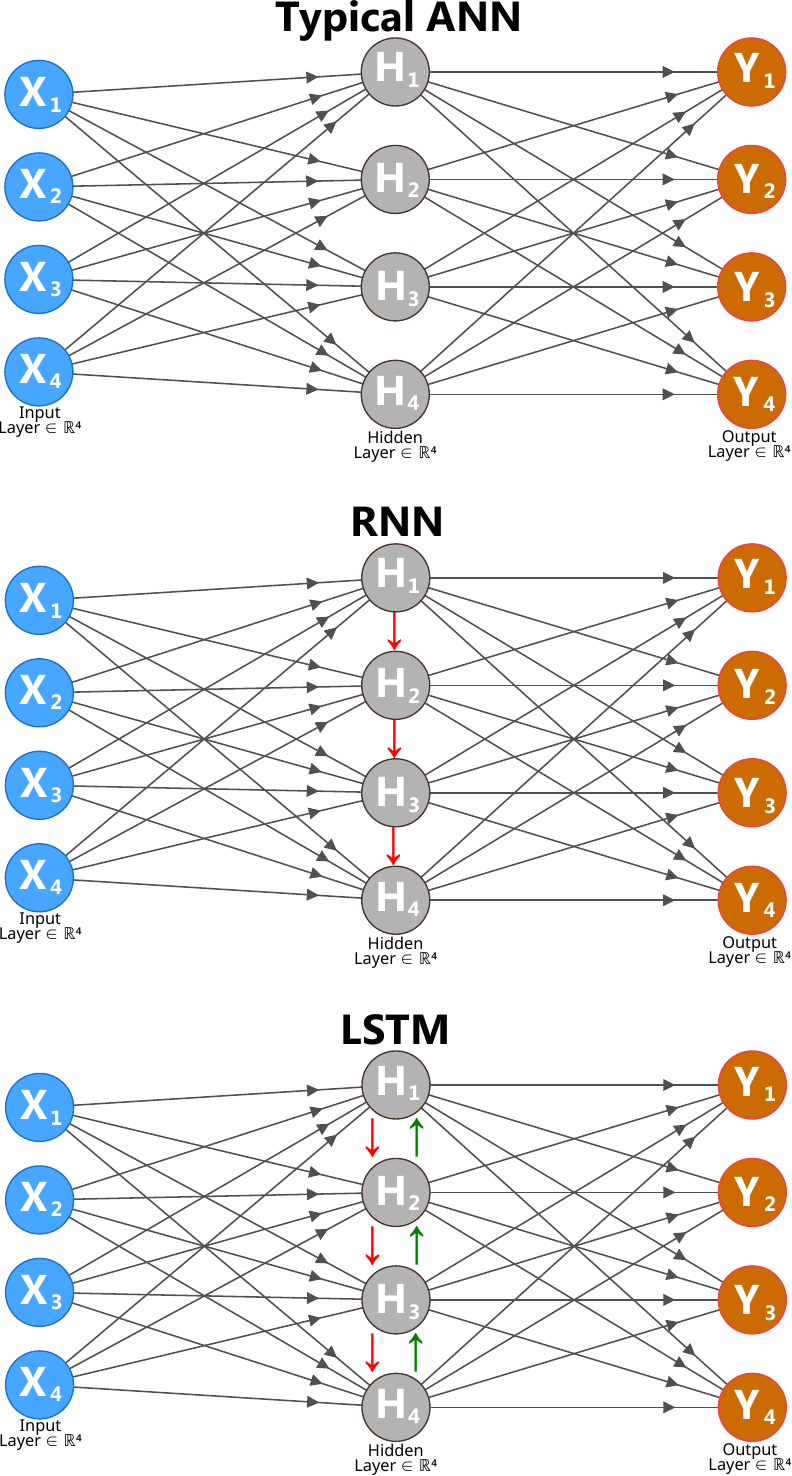}
    \caption{Comparison of ANN, RNN and LSTM architectures.
             The figure copied from~\cite{ma2019spatiotemporal}.}
    \label{fig:rnn1}
\end{figure}

Recurrent Neural Networks (RNNs) were introduced precisely for this
purpose. Unlike standard ANNs, an RNN cell maintains a hidden state
that is updated at each time step, enabling information to persist
across the sequence. Figure~\ref{fig:rnn1} contrasts a typical ANN,
an RNN, and an LSTM in terms of their connectivity
patterns~\cite{ma2019spatiotemporal}.

Despite their appeal, vanilla RNNs suffer from two well-known
limitations. First, the \emph{vanishing gradient problem}: as the
distance between two time steps increases, the gradient signal used
during back-propagation through time diminishes exponentially,
making it difficult for the network to learn long-range dependencies.
Second, the sliding-window nature of RNNs means that context outside
the window is entirely
discarded~\cite{Nguyen_1, Nguyen_2, olah2015understanding}.

\begin{figure}[htbp]
    \centering
    \includegraphics[width=0.9\linewidth]{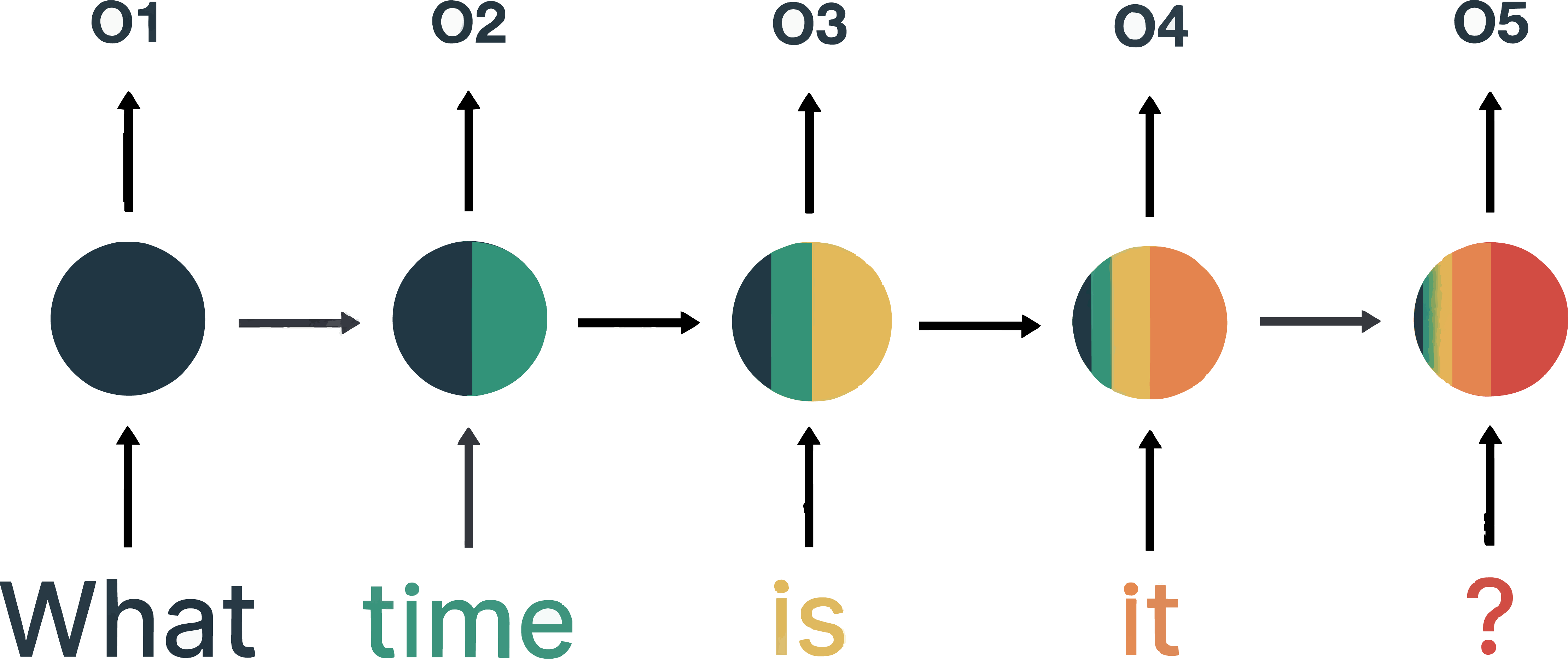}
    \caption{Sentence processing scenario illustrating decreasing
             interaction with distance.
             The figure copied from~\cite{Nguyen_1}.}
    \label{fig:rnn2}
\end{figure}

Figure~\ref{fig:rnn2} illustrates how local context dominates in an
RNN: adjacent tokens interact strongly, while distant tokens interact
weakly. Real-world sequences, including network flows, often contain
dependencies that span many time steps, which plain RNNs cannot
model reliably~\cite{Nguyen_1, Nguyen_2}.

\begin{figure}[htbp]
    \centering
    \includegraphics[width=\linewidth]{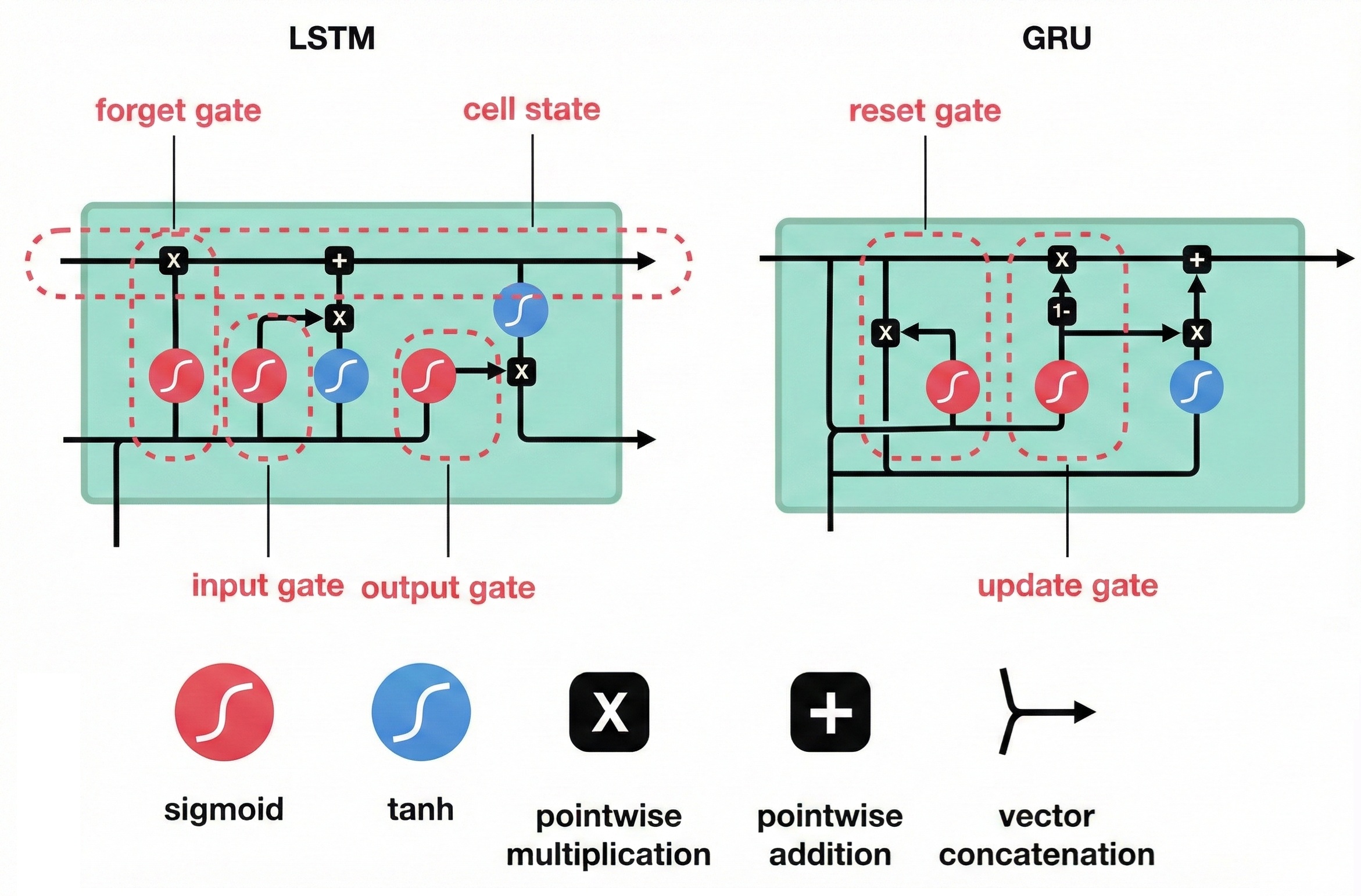}
    \caption{LSTM and GRU architectures.
             The figure copied from~\cite{Nguyen_2}.}
    \label{fig:rnn3}
\end{figure}

Long Short-Term Memory (LSTM)~\cite{olah2015understanding} and Gated
Recurrent Unit (GRU)~\cite{Nguyen_2} were developed to overcome
these shortcomings. Both architectures introduce gating mechanisms
that allow the network to selectively retain or discard information
at each time step, regardless of its position in the sequence.
Important information is propagated forward; unimportant information
is forgotten. The internal structures of LSTM and GRU are depicted
in Figure~\ref{fig:rnn3}.

Network traffic analysis is a natural application domain for
RNN-family models, because a network flow is precisely a time series
of packets governed by strict protocol rules. Numerous studies have
exploited this structure for IoT security
tasks~\cite{haddadpajouh2018deep, koroniotis2019towards,
moustafa2018ensemble, ortiz2019devicemien, lopez2017network}.
Among these, the work of Lopez-Martin et al.~\cite{lopez2017network}
is particularly relevant: they apply a CNN--RNN hybrid to classify
IoT network traffic, using the first 20 TCP/UDP packets per flow
(six features each) to construct a $6\times20$ time-series matrix.
Our methodology follows a closely related strategy, adapted to the
Aalto university IoT devices captures dataset (Aalto dataset)~\cite{aalto2017dataset,miettinen2017iot} and extended with a systematic study of
sequence-length effects.

The remainder of this paper is organised as follows. Section~\ref{sec:methodology} describes the full pipeline from raw PCAP capture to model evaluation and  details the experimental setup and dataset characteristics. Section~\ref{sec:results} presents quantitative results and visualisations, which are then critically analysed in Section~\ref{sec:discussion}. Finally, Section~\ref{sec:conclusion} concludes the paper.

\section{Methodology}
\label{sec:methodology}

\begin{figure}[htbp]
    \centering
    \includegraphics[width=\linewidth]{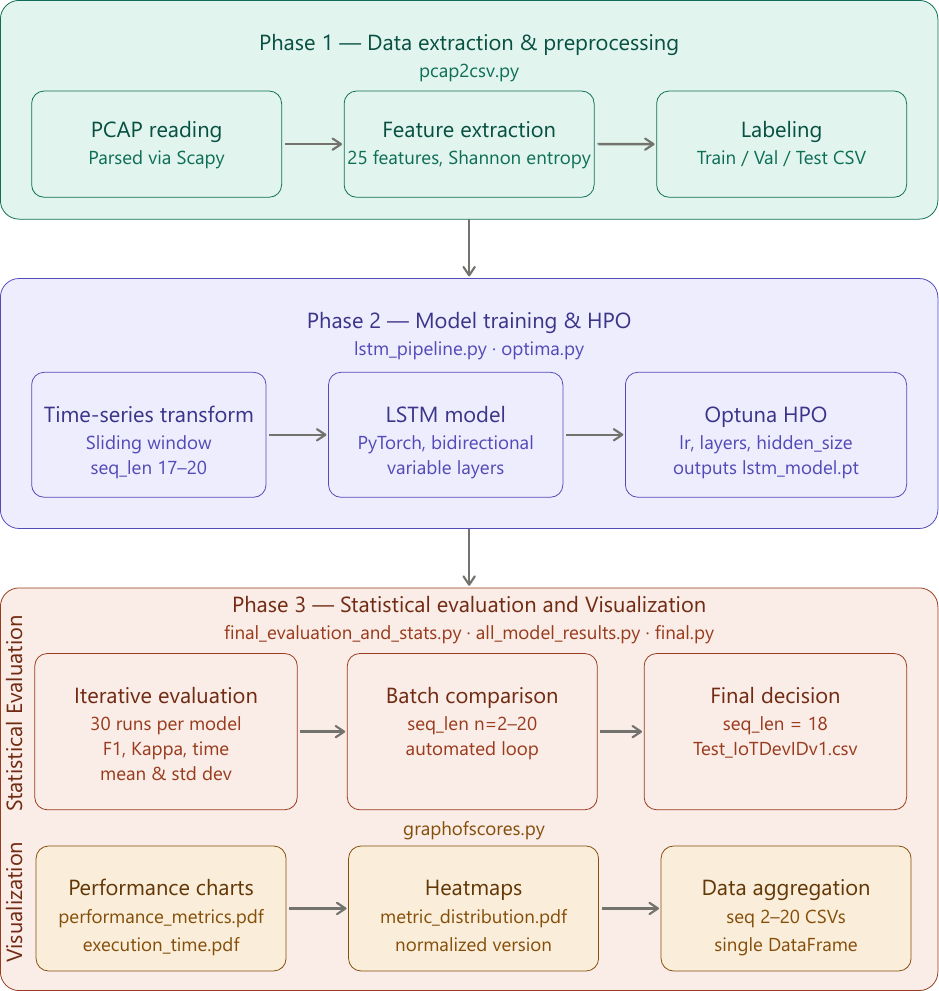}
    \caption{End-to-end pipeline: from raw PCAP captures to final
             performance graphs.}
    \label{fig:pipeline}
\end{figure}

The proposed pipeline is divided into four logical phases:
(1) data extraction and preparation,
(2) model training and hyper-parameter optimisation,
(3) Statistical evaluation, comparative analysis, and visualisation.
Figure~\ref{fig:pipeline} summarises the overall workflow.

\subsection{Phase 1 – Data Extraction and Preparation}
\label{sec:phase1}

Raw network traffic is captured in PCAP (Packet Capture)
format from the Aalto dataset~\cite{aalto2017dataset,miettinen2017iot}. Each capture file is
parsed using the \emph{Scapy} library, and \textbf{25 features} per packet
are extracted according to OSI model layers. These features are not
arbitrarily selected; rather, they are based on the original feature
set proposed in the first version of the IoTDeVID
study~\cite{kostas2021iotdevid1}.

\begin{itemize}
    \item \textbf{Protocol flags (Layers 2–7):} binary indicators
          for ARP, LLC, IP, ICMP, TCP, UDP, HTTP, DNS, DHCP, and
          related protocols.
    \item \textbf{Packet metrics:} raw packet size and port-class
          labels for source and destination ports.
    \item \textbf{Payload entropy:} Shannon entropy of the packet
          payload, providing a scalar measure of data complexity.
\end{itemize}

Device labels are derived by matching MAC addresses against a
pre-defined lookup table of 27 known IoT device types. After feature
extraction, packets are assigned to the pre-defined Train, Validation,
and Test splits for downstream processing.

\subsection{Phase 2 – Model Training and Hyper-parameter Optimisation}
\label{sec:phase2}

\subsubsection{Time-Series Conversion}
The tabular feature data are standardised via z-score normalisation and
integer-encoded for class labels. A \emph{sliding window} of length
$\ell$ is then applied to transform the flat feature vectors into
ordered tensors of shape $(\ell \times 25)$, one tensor per sample.

\subsubsection{LSTM Architecture}
The classifier is implemented in PyTorch as a configurable LSTM-based
neural network. Key architectural choices include:
\begin{itemize}
    \item \textbf{Bidirectionality:} optionally enabled, doubling the
          effective hidden state by processing the sequence both
          forwards and backwards.
    \item \textbf{Number of stacked LSTM layers:} searched during
          hyper-parameter optimisation.
    \item \textbf{Hidden size:} the dimensionality of the LSTM hidden
          state, also subject to optimisation.
\end{itemize}
Class imbalance is addressed by computing inverse-frequency class
weights and passing them to the cross-entropy loss function.

\subsubsection{Hyper-parameter Optimisation (HPO)}
The \emph{Optuna} framework~\cite{akiba2019optuna} is used for
automated HPO. The search space covers:
\begin{itemize}
    \item Learning rate,
    \item Number of stacked layers,
    \item Hidden unit count,
    \item Bidirectionality (enabled or disabled).
\end{itemize}
The optimisation procedure iterates over a range of sequence lengths
($\ell \in \{2, 3, \ldots, 20\}$), invoking the full HPO pipeline for
each value and retaining the best model checkpoint alongside
normalised confusion matrices and learning-curve plots.

\subsection{Phase 3 – Statistical Evaluation, Comparative Analysis, and Visualisation}
\label{sec:phase3}

\subsubsection{Iterative Evaluation}
To obtain statistically reliable estimates, the saved model is
reloaded and evaluated 30 independent times on the held-out test set.
Each run records Accuracy, Balanced Accuracy, Precision, Recall,
F1-Score, Cohen's Kappa, and execution time.

\subsubsection{Batch Comparison Across Sequence Lengths}
The evaluation described above is automated for every sequence length
$\ell \in \{2, 3, \ldots, 20\}$, generating per-length summary
statistics containing means and standard deviations across the 30
iterations. These aggregated statistics are subsequently used to
visualise the relationship between sequence length and model
performance through line charts (Figure~\ref{fig:line}) and metric heatmaps
(Figure~\ref{fig:heatmap}).

\subsubsection{Final Decision}
Based on the batch comparison and the generated visual analyses,
sequence length $\ell = 18$ is selected as a robust optimal
trade-off. The definitive evaluation is then executed on the main
test set using this configuration, and the final performance figures
are reported.

\subsection{Experimental Setup}
\label{sec:setup}

The dataset contains 540 sessions drawn from 27 IoT device classes
(20 sessions per classes). The sessions are split into fixed Train, Validation, and Test
subsets. Each session is converted into a sequence of packets; no
inter-session overlap is permitted so that the model cannot exploit
cross-session leakage.

All experiments are conducted with the same random seed to ensure
reproducibility. The Optuna HPO budget is 100 trials per
sequence-length value, optimising macro-averaged F1-score on the
validation set.

\section{Results}
\label{sec:results}

\subsection{Effect of Sequence Length on Performance}

\begin{figure}[htbp]
    \centering
    \includegraphics[width=\linewidth]{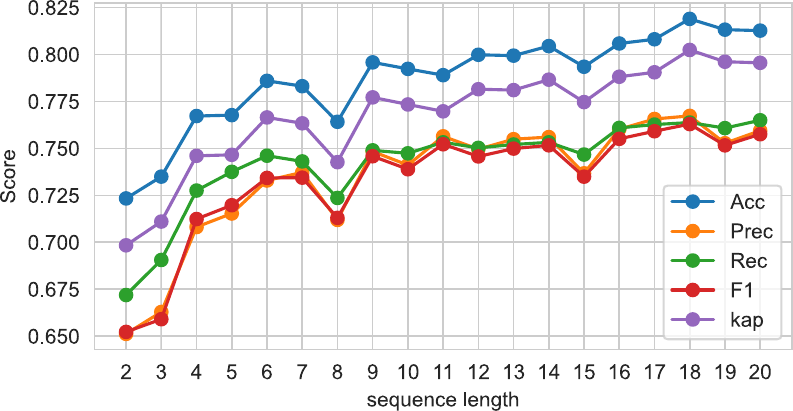}
    \caption{Variation of performance metrics with respect to
             sequence length.}
    \label{fig:line}
\end{figure}

Figure~\ref{fig:line} shows how the five main performance metrics
evolve as sequence length increases from 2 to 20.

Performance rises approximately linearly from $\ell = 2$ to
$\ell = 6$, reflecting the model's increasing ability to exploit
contextual packet information. Beyond $\ell = 6$, the improvement
pattern becomes wave-like, with local maxima and minima. The highest
aggregate performance is observed at $\ell = 18$.

\begin{figure}[htbp]
    \centering
    \includegraphics[width=\linewidth]{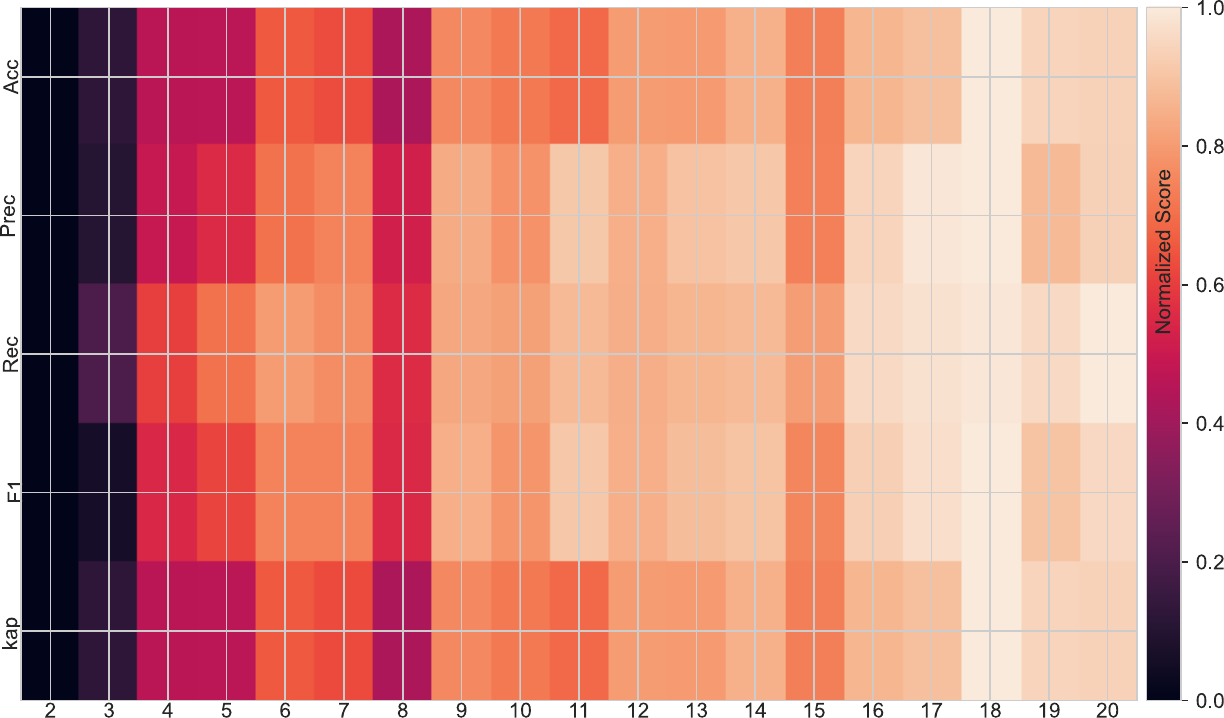}
    \caption{Heatmap representation of performance metric values
             across different sequence lengths.}
    \label{fig:heatmap}
\end{figure}

Figure~\ref{fig:heatmap} provides a complementary heatmap view,
making it easy to compare metric behaviour across the full range of
sequence lengths simultaneously.

\subsection{Confusion Matrix}

\begin{figure}[htbp]
    \centering
    \includegraphics[width=\linewidth]{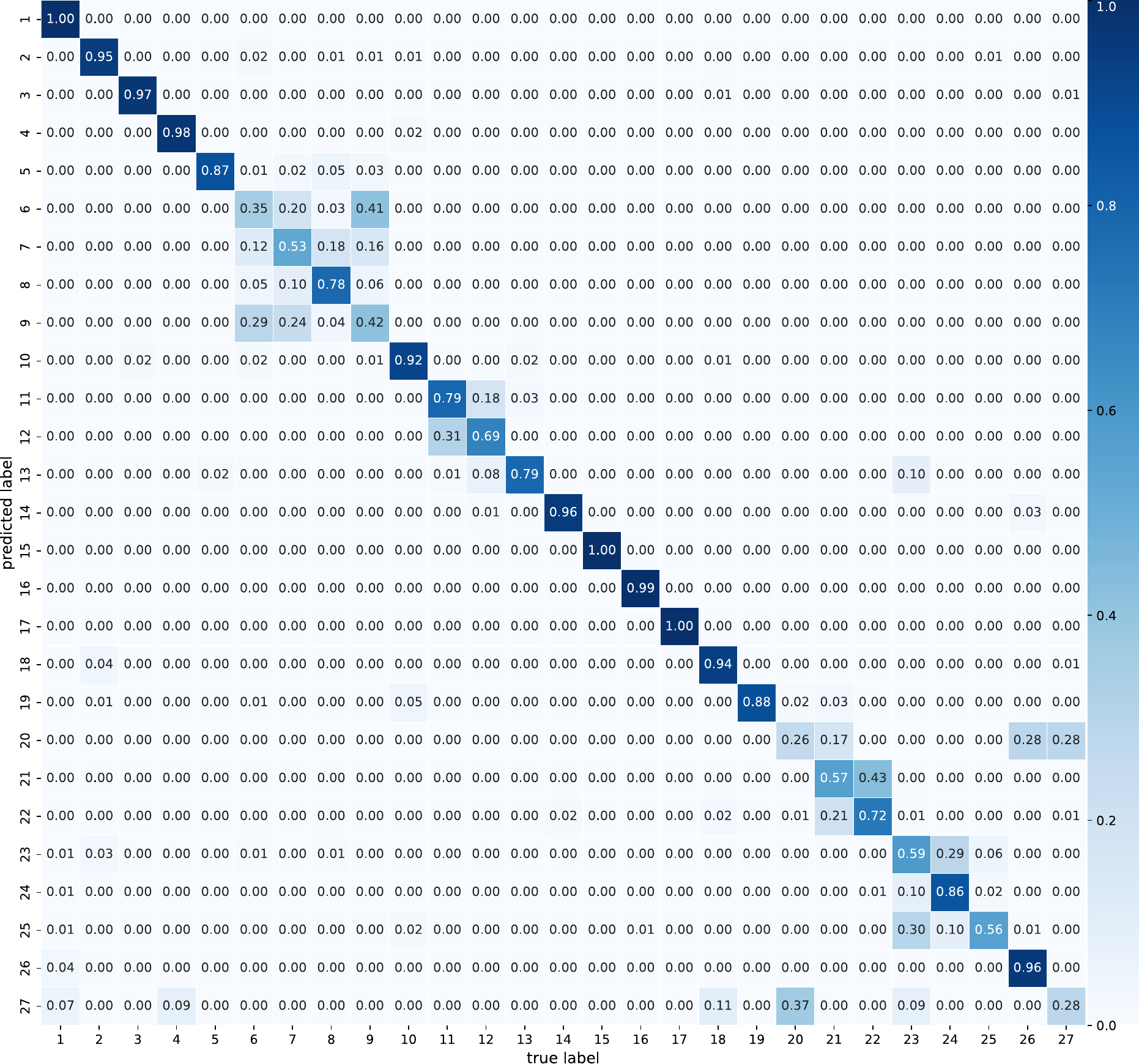}
    \caption{Normalised confusion matrix for the final model
             ($\ell = 18$) on the held-out test set.
             Rows are true labels; columns are predicted
             labels. See Table~\ref{tab:device_scores}
for device names.}
    \label{fig:cm}
\end{figure}

Figure~\ref{fig:cm} presents the normalised confusion matrix obtained
with $\ell = 18$ on the held-out test set. The diagonal entries
indicate per-class recall. Most device classes achieve recall above
0.87; the notable exceptions are the D-Link sensor cluster
(D-LinkSensor, D-LinkSiren, D-LinkWaterSensor), where mutual
confusion is caused by highly similar protocol fingerprints, and the
low-sample classes SmarterCoffee and iKettle2.

\subsection{Per-Device Classification Performance}

Table~\ref{tab:device_scores} reports Precision, Recall, F1-Score,
and Support for each of the 27 device classes. Devices with
distinctive protocol behaviour (HomeMaticPlug, HueBridge,
HueSwitch, MAXGateway) achieve near-perfect scores. Devices that
share communication patterns or belong to the same manufacturer
family (D-Link sensor group, TP-LinkPlug variants) show lower
F1-scores, highlighting a fundamental challenge in network-based IoT
fingerprinting.

\begin{table}[htbp]
    \centering
    \caption{Device-Level Classification Scores ($\ell=18$)}
    \label{tab:device_scores}
    \begin{tabular}{clrrrr}
        \toprule
        \textbf{No} & \textbf{Device} & \textbf{Prec.} & \textbf{Rec.} &
        \textbf{F1} & \textbf{Support} \\
        \midrule
        1  & Aria                & 0.75 & 1.00 & 0.86 & 113 \\
        2  & D-LinkCam           & 0.91 & 0.95 & 0.93 & 1236 \\
        3  & D-LinkDayCam        & 0.98 & 0.97 & 0.98 & 305 \\
        4  & D-LinkDoorSensor    & 0.98 & 0.98 & 0.98 & 515 \\
        5  & D-LinkHomeHub       & 0.98 & 0.87 & 0.92 & 1851 \\
        6  & D-LinkSensor        & 0.41 & 0.35 & 0.38 & 1573 \\
        7  & D-LinkSiren         & 0.47 & 0.53 & 0.50 & 1518 \\
        8  & D-LinkSwitch        & 0.71 & 0.78 & 0.74 & 1630 \\
        9  & D-LinkWaterSensor   & 0.39 & 0.42 & 0.40 & 1618 \\
        10 & EdimaxCam           & 0.80 & 0.92 & 0.86 & 235 \\
        11 & EdimaxPlug1101W     & 0.69 & 0.79 & 0.73 & 275 \\
        12 & EdimaxPlug2101W     & 0.77 & 0.69 & 0.72 & 299 \\
        13 & EdnetCam            & 0.85 & 0.79 & 0.82 & 90 \\
        14 & EdnetGateway        & 0.97 & 0.96 & 0.96 & 203 \\
        15 & HomeMaticPlug       & 1.00 & 1.00 & 1.00 & 285 \\
        16 & HueBridge           & 1.00 & 0.99 & 0.99 & 3763 \\
        17 & HueSwitch           & 1.00 & 1.00 & 1.00 & 4374 \\
        18 & Lightify            & 0.98 & 0.94 & 0.96 & 1171 \\
        19 & MAXGateway          & 1.00 & 0.88 & 0.94 & 145 \\
        20 & SmarterCoffee       & 0.32 & 0.26 & 0.29 & 46 \\
        21 & TP-LinkPlugHS100    & 0.66 & 0.57 & 0.61 & 189 \\
        22 & TP-LinkPlugHS110    & 0.55 & 0.72 & 0.63 & 169 \\
        23 & WeMoInsightSwitch   & 0.66 & 0.59 & 0.62 & 1703 \\
        24 & WeMoLink            & 0.70 & 0.86 & 0.77 & 1609 \\
        25 & WeMoSwitch          & 0.83 & 0.56 & 0.67 & 1120 \\
        26 & Withings            & 0.87 & 0.96 & 0.92 & 194 \\
        27 & iKettle2            & 0.28 & 0.28 & 0.28 & 46 \\
        \midrule
        -- & \textbf{Mean}       & 0.76 & 0.76 & 0.76 & -- \\
        \bottomrule
    \end{tabular}
\end{table}

\subsection{Summary Statistics}

Table~\ref{tab:summary_stats} provides the final aggregate statistics
computed over 30 repeated inference runs on the test set.
The negligibly small standard deviations for all metrics except
execution time confirm that the evaluation is fully deterministic
once the model weights are fixed; only runtime varies due
to system-level scheduling jitter.

\begin{table}[htbp]
    \centering
    \caption{Final Summary Statistics (30 Iterations, $\ell=18$)}
    \label{tab:summary_stats}
    \begin{tabular}{lrr}
        \toprule
        \textbf{Metric} & \textbf{Mean} & \textbf{Std} \\
        \midrule
        Accuracy            & 0.7985 & $3.39 \times 10^{-16}$ \\
        Balanced Accuracy   & 0.7631 & $1.13 \times 10^{-16}$ \\
        Precision           & 0.7583 & $2.26 \times 10^{-16}$ \\
        Recall              & 0.7631 & $1.13 \times 10^{-16}$ \\
        F1-Score            & 0.7570 & $2.26 \times 10^{-16}$ \\
        Cohen's Kappa       & 0.7803 & $2.26 \times 10^{-16}$ \\
        Execution Time (s)  & 1.2396 & 0.6060 \\
        \bottomrule
    \end{tabular}
\end{table}

\section{Discussion}
\label{sec:discussion}

The results demonstrate that LSTM-based models can effectively
fingerprint IoT devices from raw network traffic with no payload
inspection (i.e., no deep packet inspection beyond header fields and
payload entropy). Several observations merit further discussion.

\textbf{Sequence length matters but saturates.}
The approximately linear gain from $\ell = 2$ to $\ell = 6$ confirms
that even a short temporal context dramatically improves
discrimination. The subsequent wave-like plateau suggests that
additional history beyond six packets provides diminishing returns
on average, though specific device pairs may benefit from longer
windows.

\textbf{D-Link sensor confusion.}
D-LinkSensor, D-LinkSiren, and D-LinkWaterSensor consistently
confuse the model. These three devices share the same manufacturer,
use overlapping port ranges, and generate packets with similar size
distributions. Distinguishing them may require application-layer
features or longer sequences that capture device-specific periodic
behaviour.

\textbf{Low-sample classes.}
SmarterCoffee and iKettle2 each contribute only 46 test samples.
Despite class-weighted training, the model struggles with these
classes, highlighting the sensitivity of the approach to support
size. Data augmentation or few-shot learning techniques could
improve performance on rare device types.

\textbf{Reproducibility.}
The near-zero standard deviations across metrics in
Table~\ref{tab:summary_stats} confirm fully deterministic inference,
which is a desirable property in security-critical deployments where
stable predictions are required.

\section{Conclusion}
\label{sec:conclusion}

We have presented an end-to-end LSTM pipeline for IoT device
identification from network packet captures. The pipeline extracts
25 packet-level features---including protocol flags, packet size,
port classifications, and Shannon entropy---converts them into
sliding-window time-series tensors, trains a bidirectional LSTM
classifier with Optuna-based HPO, and rigorously evaluates models
across sequence lengths 2–20.

On the Aalto dataset containing 27 device classes, the
optimal configuration ($\ell = 18$) achieves \textbf{79.85\% accuracy}
and a macro F1-score of \textbf{75.70\%}. Performance is excellent
for devices with distinctive network fingerprints and degrades for
functionally similar devices within the same product family,
pointing toward the need for richer feature representations or
complementary identification signals in future work.

The complete source code is publicly available at:
\url{https://github.com/kahramankostas/LSTM-based-IoT-Device-Identification}.


	\bibliographystyle{ieeetr}
	\bibliography{simple}
	
\end{document}